\documentclass[useAMS,usenatbib]{mn2e}
\usepackage{epsfig}
\usepackage{textcomp}

\def\mum{\textmu m}
\def\mums{\textmu m }

\title[Morphological Studies of the SWIRE Galaxy Population]
{Morphological Studies of the SWIRE Galaxy Population in the UGC 10214 HST/ACS field}

\author[Hatziminaoglou et al.]{E. Hatziminaoglou$^{1}$, P. Cassata$^{2}$, G. Rodighiero$^{2}$,
I. P\'erez-Fournon$^{1}$,
\newauthor A. Franceschini$^{2}$, A. Hern\'{a}n-Caballero$^{1}$, F.M. Montenegro-Montes$^{1}$
\newauthor A. Afonso-Luis$^{1}$, T. Jarrett$^{3}$, G. Stacey$^{4}$, C. Lonsdale$^{5}$, F. Fang$^{4}$, S. Oliver$^{6}$,
\newauthor M. Rowan-Robinson$^{7}$, D. Shupe$^{3}$, H.E. Smith$^{5}$, J. Surace$^{3}$, C.K. Xu$^{3}$,
\newauthor E.A. Gonz\'{a}lez-Solares$^{8}$\\
$^{1}$Institute de Astrofisica de Canarias, C/ Via Lactea s/n, E-38200 La Laguna, Spain\\
$^{2}$Dipartimento di Astronomia, Universita di Padova, Vicolo Osservatorio 5, 35122 Padua, Italy\\
$^{3}$Infrared Processing and Analysis Center, California Institute of Technology, Pasadena, CA 91125, USA\\
$^{4}$Cornell University, Astronomy Department, Ithaca, NY 14853, USA\\
$^{5}$Center for Astrophysics and Space Sciences, University of California, San Diego, La Jolla,
CA 92093-0424, USA\\
$^{6}$Astronomy Centre, Department of Physics and Astronomy, University of Sussex, Falmer,
Brighton BN1 9QJ, UK\\
$^{7}$Astrophysics Group, Blackett Laboratory, Imperial College London, London SW7 2BW, UK\\
$^{8}$Institute of Astronomy, University of Cambridge, Madingley Road, Cambridge CB3 0HA, UK}

\begin{document}

\maketitle

\begin{abstract}

We present results of a morphological analysis of a small subset of the Spitzer 
Wide-area InfraRed Extragalactic survey (SWIRE) galaxy population. The analysis 
is based on public ACS data taken inside the SWIRE N1 field, which are the deepest 
optical high-resolution imaging available within the SWIRE fields as of today. Our 
reference sample includes 156 galaxies detected by both ACS and SWIRE. Among 
the various galaxy morphologies, we disentangle two main classes, spheroids (or 
bulge-dominated galaxies) and disk-dominated ones, for which we compute the number 
counts as a function of flux. We then limit our sample to objects with IRAC fluxes 
brighter than 10 \textmu Jy, estimated $\sim$90\% completeness limit of the SWIRE 
catalogues, and compare the observed counts to model predictions. We find that the 
observed counts of the spheroidal population agree with the expectations of a 
hierarchical model while a monolithic scenario predicts steeper counts. Both scenaria, 
however, under-predict the number of late-type galaxies. These observations show 
that the large majority (close to 80 per cent) of the 3.6 and 4.5 \mums galaxy 
population, even at these moderately faint fluxes, is dominated by spiral and 
irregular galaxies or mergers.

\end{abstract}

\maketitle

\section{INTRODUCTION}
\label{intro}
A key question on galaxy formation and evolution, in particular about the early-type 
sub-population which includes the most massive galaxies at any redshifts, is when 
and on which timescales their stellar content has been formed and assembled. 
Two schematic models are often confronted with the observations, the monolithic 
collapse model (\citealt{eggen62}; \citealt{larson75}; \citealt{chiosi02})
and the hierarchical assembly scenario (e.g. \citealt{white78};
\citealt{white91}; \citealt{somerville99}; \citealt{cole00}).
In the monolithic collapse, a burst of star formation happened at very high redshifts 
($z_{form} \ge$3) and was followed by passive evolution of the stellar populations, 
whereas in the hierarchical assembly scenario the timescales for more massive 
galaxies are longer, resulting in somewhat younger mean ages. According to the latter
interpretation, ellipticals are formed by mergers and/or accretion of smaller 
galaxies over timescales comparable to the Hubble time (see e.g. \citealt{bell04}; \citealt{faber05}).

Fundamental observational constraints on the star formation history and the formation pattern
are provided by the broad-band colours, line strength indexes and stellar chemical 
abundances. When referred to massive ellipticals, these data suggest that the bulk of 
stars might have been formed in a remote past. However, some secondary activity of 
star formation in the recent past is also evident: nearby ellipticals show a large 
variety of morphological and kinematic peculiarities (e.g. \citealt{longhetti00})
and a substantial spread of stellar ages, particularly for the field population 
\citep{thomas05}. 
Strong evolution in the population of early-type galaxies has been 
reported by \cite{kauffmann96} and \cite{kauffmann98}
which has been considered to support the hierarchical galaxy formation models.

The current results about the number counts and redshift distributions 
of the evolved galaxies at high redshift are still rather inconclusive, and theoretical
models about their formation and evolution are correspondingly uncertain
(see \citealt{somerville04} for a brief review).

Near-infrared surveys are best suited for the study of faint high-redshift galaxy 
populations, for various reasons: the observed fluxes are minimally influenced 
by K-corrections and only weakly affected by dust extinction, and at the same time
good indicators of the stellar mass content of galaxies \citep{dickinson03}.  
Mid-infrared observations, on their side, are strongly informative about phases of 
active star formation in galaxies (\citealt{franceschini01}; \citealt{rowan01}; 
\citealt{xu03} and many others).
After the ISO exploratory observations, the Spitzer Space Observatory mission 
has started to map systematically the high-redshift universe at such long wavelengths.

The main aim of the Spitzer Wide-area InfraRed Extragalactic survey (SWIRE) is 
the study of the evolution of both actively star-forming and
passively-evolving galaxies, by exploiting a huge sample of some 2 million galaxies
that are detected within the $\sim$ 50 deg$^2$ of the survey.

Good quality optical data are essential for the identification and characterization 
of Spitzer sources. However, typical images obtainable from the ground are not adequate 
for accurate morphological analysis and source classification.
In this paper we exploit very deep multiband public images taken with the Advanced Camera 
for Surveys (hereafter ACS; \citealt{ford98}) in a small portion
of the SWIRE N1 field, in order to study the morphological properties of the 
infrared (IR) emitting galaxies. In what follows and unless specified otherwise, 
IR refers to 3.6 and 4.5 \mum.

Section \ref{data} of the paper describes the optical (HST/ACS) and IR (Spitzer) 
data used in our analysis. Section \ref{tools} gives a brief description of the tools 
used for the morphological analysis and classification as well as a comparison of 
their results. Section \ref{numbers} describes a comparison of our modelistic galaxy 
number counts with our observational data. Finally, sections \ref{discuss} and
\ref{conclude} summarise our results and conclusions.

\section{THE DATA}
\label{data}

SWIRE N1 was the first field to be observed by Spitzer for the SWIRE Legacy program.
The IR data used here were obtained in February 2004 and the SWIRE catalogues we use
throughout this work were processed by the SWIRE team.
Source extraction was performed using {\it SExtractor}
\citep{bertin96} with local background mesh subtraction. Kron and 2.9" aperture
fluxes were used here, for extended and point-like sources, respectively.
The Kron fluxes were extracted within a minimum radius of 2" using a Kron factor
of 2.5. These fluxes were the {\it default} fluxes in the first SWIRE data release.
The 3.6 and 4.5 \mums counts are roughly 90\% complete at around 10 \textmu Jy.
The astrometric accuracy is better than 1".
More details about the entire data set can be found in \citealt{lonsdale04}, 
Surace et al. (in preparation) and Shupe et al. (in preparation), 
as well as the data release document \citep{dr1}.

The optical data were obtained with the ACS on the Hubble Space Telescope.
The field is centred on UGC 10214 (also known as VV 29 and Arp 188, the 
{\sl Tadpole} galaxy), a bright spiral at $z=0.032$ with a huge tidal tail. 
The optical catalogue contains some 5700 objects with 10 $\sigma$
detection limits for point sources of of $27.8$, $27.6$ and $27.2$
AB magnitudes in the $g_{\rm F475W}$, $V_{\rm F606W}$ and $I_{\rm F814W}$
bands respectively. For details on the data reduction and catalogue production see
\citealt{benitez04}. This field lies close to the centre of the SWIRE N1 field, 
providing an unprecedented opportunity of combining very deep optical data with 
high quality IR data in wavelength ranges so far poorly explored.

Within the $\sim$14 sq. arcmin ACS field, we use an area of $\sim$10 sq. arcmin
well outside the region occupied by the UGC 10214 galaxy.
Any remaining background contamination likely to affect the IRAC fluxes 
is accounted for thanks to the way the SWIRE catalogues
are created: they use a locally generated background estimate and any 
underlying faint extended emission is considered as background and subtracted.

The matching between the SWIRE and ACS sources was done using the {\sc tmatch}
task of {\it IRAF}, with an 1" match radius. The sources were then visualised one by
one and a flag was assigned to each one of them, stating their status with 
respect to close counterparts. Out of the 177 matched objects, 175 and 130 are
detected in 3.6 and 4.5 \mums, respectively (with two having only
4.5 \mums detections) and 18 (i.e. $\sim 10$\%) 
have at least two extracted ACS counterparts within an 1" radius.
For six of them, however, we can clearly identify the object that mostly contributes 
to the IRAC fluxes, as the difference in I-band magnitude between the counterparts is 
larger than 2.5.
For a detailed discussion of the optical/IR bandmerged catalogues and the
spectral energy distribution (SED) of the matched objects we defer to
P\'{e}rez-Fournon et al. (in preparation) and Hern\'{a}n-Caballero
et al. (in preparation).

\begin{figure*}
\centerline{
\psfig{file=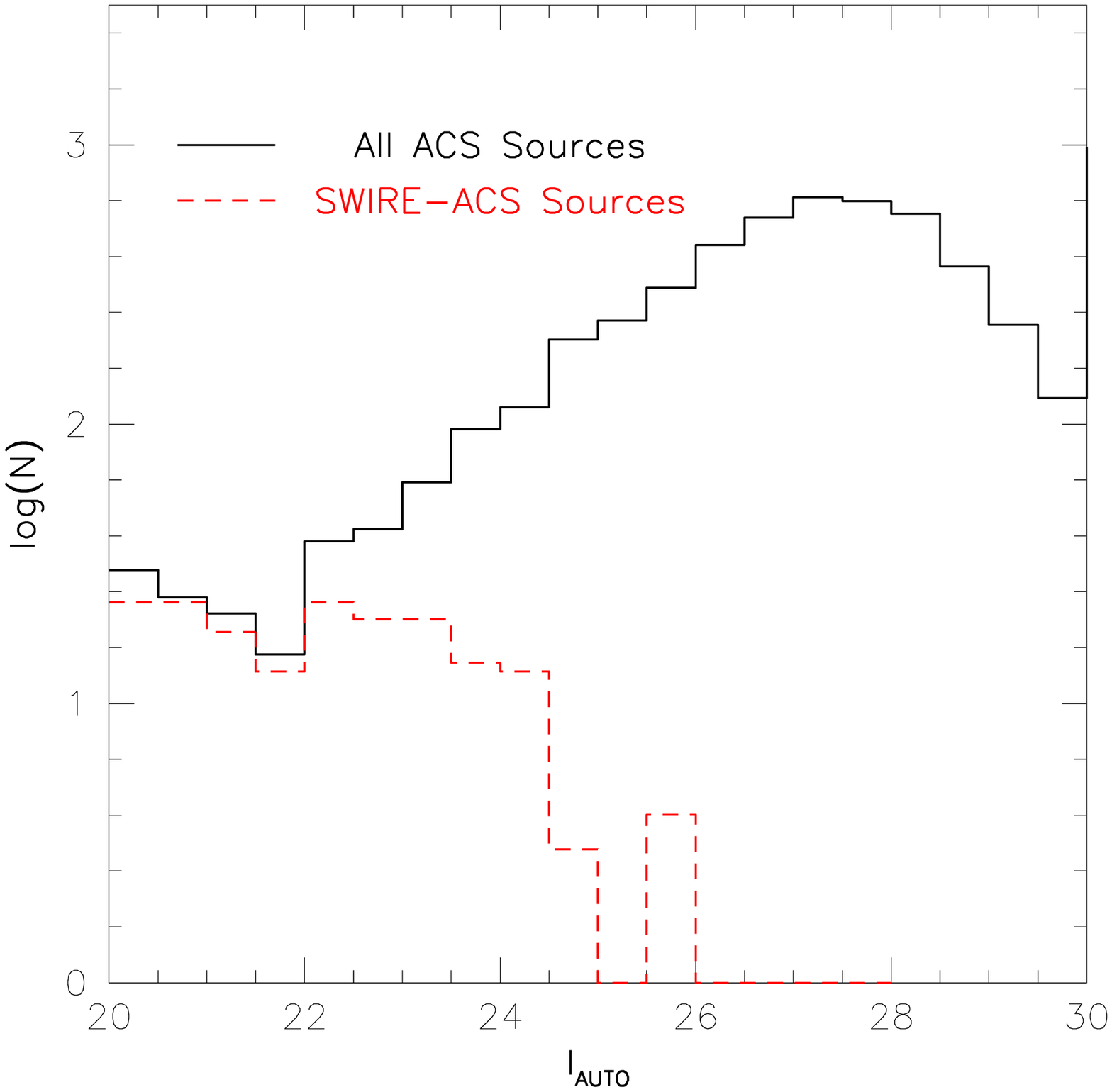,width=8cm}
\psfig{file=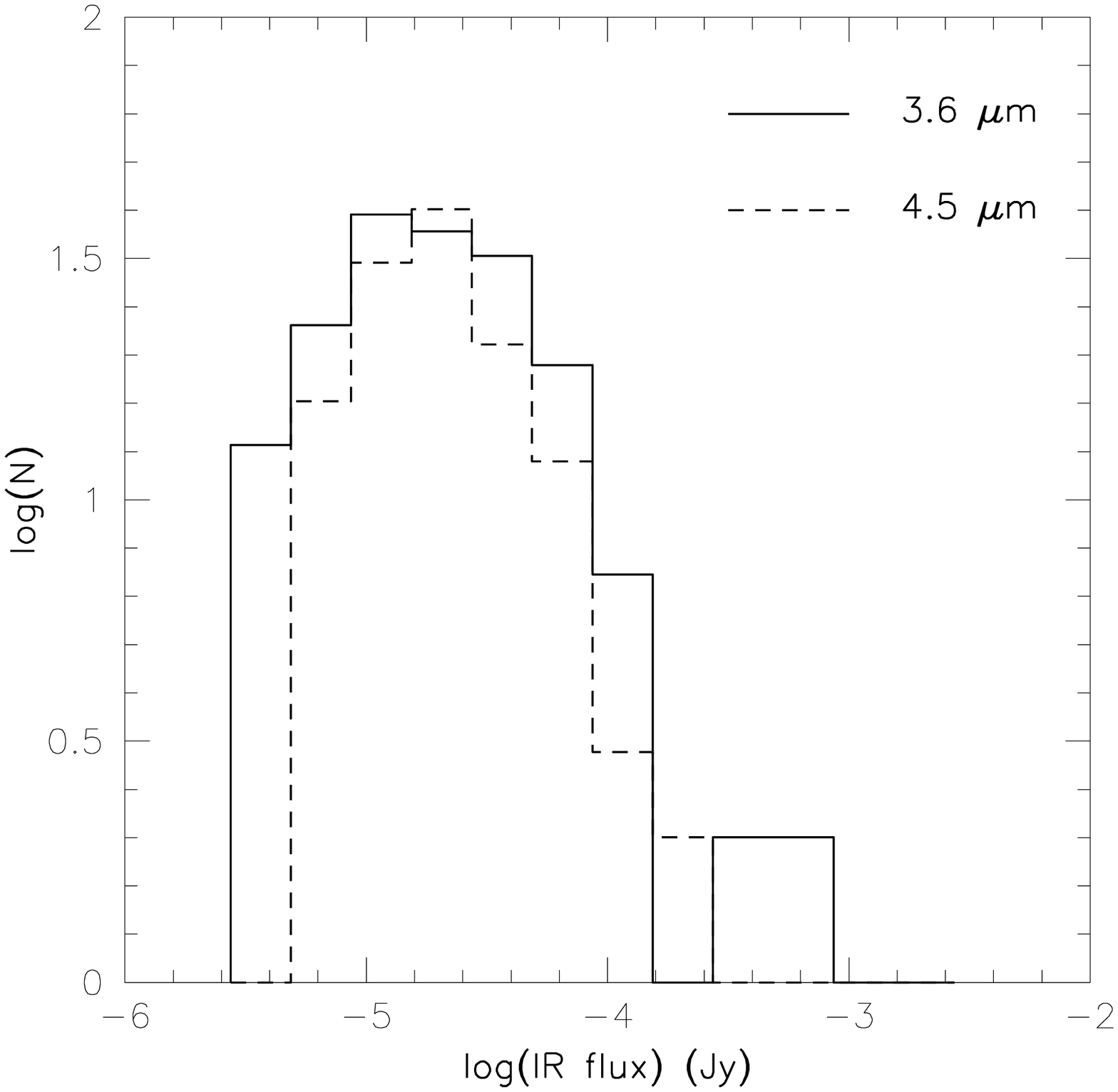,width=8cm}}
\caption{Left panel: I-band number counts for all ACS sources (solid line) and for the SWIRE-ACS
identifications (dashed line) in the $\sim$ 10 sq. arcmin field covered by the
ACS observations, after the area around UGC 10214 has been removed. Right panel: 3.6 \mums (solid
line) and 4.5 \mums (dashed line) distribution of the SWIRE-ACS sources.}
\label{figInumbercounts}
\end{figure*}

\begin{figure}
\centerline{
\psfig{file=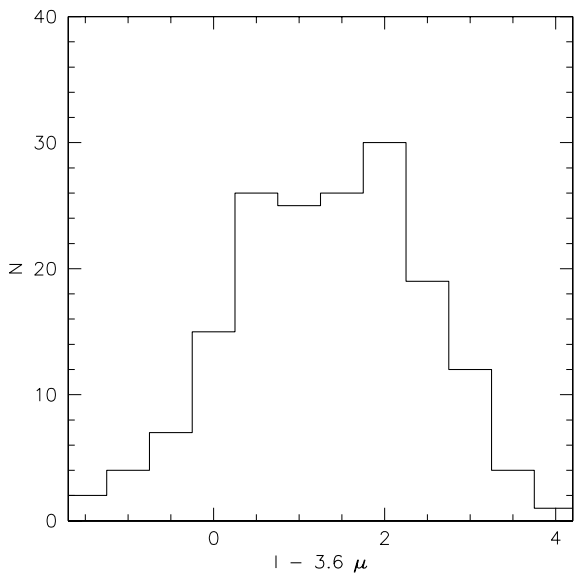,width=8cm}}
\caption{(I-3.6\mum)$_{AB}$ colour histogram for the matched SWIRE-ACS sources.}
\label{figcolhisto}
\end{figure}

The I-band counts for the entire ACS field and for the sources with a SWIRE
counterpart are shown in Fig. \ref{figInumbercounts} in solid and
in dashed lines, respectively. Down to a magnitude of I$\sim$ 22.5
almost all ACS sources have an IR counterpart, while the number starts 
dropping below I$\sim$ 23.5, reflecting the shallowness of the SWIRE data.
The completeness of the 3.6 \mums counts drops quickly below 10 \textmu Jy,
value corresponding to an AB limiting magnitude of 21.4 or I$\sim$ 23.5 for
a typical I-3.6\mums colour of 2 (see Fig. \ref{figcolhisto} for the colour
distribution of the 177 matched sources).

Note that the probability of chance associations within 1" of radius is of less
than 2\% up to an I-band magnitude of 24.0 and up to 5\% at an I-band magnitude of 25.
Taking into account the I-band magnitude distribution of the SWIRE sources 
(Fig. \ref{figInumbercounts}), one expects less than 5 out of the 177 sources to 
be wrongly matched.

\section{MORPHOLOGICAL CLASSIFICATION}
\label{morph}

\subsection{Tools for the morphological analysis}
\label{tools}

A number of parametric and non-parametric methods are suggested in the literature
for quantitative assessment of the morphological properties of intermediate and high 
redshift galaxies. Among the parametric approaches, an often used representation of the 
galaxy morphological types is the S\'ersic index, $n$, which appears in the
surface brightness profile law $I_r = I_b(0)e^{-b_n(r/r_m)^{1/n}}$, where $I_b(0)$ is 
the bulge central intensity, $r_m$ is the bulge semi-major effective radius and $n$ 
the S\'ersic shape parameter \citep{sersic68}. The quantity $b_n$ is a function of $n$, 
and is chosen so that $r_m$ encloses
half of the total luminosity. The S\'ersic profile includes the classical de Vaucouleurs
profile when $n$ is equal to 4.

Alternatively, non-parametric approaches have been proposed,
with the most common one using the combination of the light Concentration ($C$),
Asymmetry ($A$) and clumpinesS ($S$), or CAS, parameters. Concentration 
\citep{abraham96} roughly correlates with the S\'ersic index, while asymmetry
\citep{conselice00} compares the image of a source with its rotated (usually by 180$^{\, \rm o}$ )
counterpart, enabling the distinction between normal and irregular 
galaxies or merging systems. The clumpiness, \citep{conselice03},
measures the uniformity of the light distribution in a galaxy.
\cite{abraham96} and later on \citep{conselice00} showed that bulge-dominated, 
disk-dominated and merging systems occupy different but often overlapping regions
in the Asymmetry - Concentration space. \cite{conselice03} finally demonstrated 
that all galaxy types can be roughly identified by their position in the $CAS$ 
three-dimensional space.

Our morphological analysis has been performed in three independent steps: i) visual
inspection, in order to attribute each object to a given morphological class based on 
features like the presence of spiral arms and/or bars, signs of interaction, multiple
nuclei etc; ii) a non-parametric analysis of the galaxy light distribution 
using the CAS parameters (see \citealt{conselice03b} for a detailed description and
\citealt{cassata05} for their exact definition as used here);
iii) a detailed analysis of the surface brightness profiles 
using the packages GALFIT \citep{peng02} and GASPHOT \citep{pignatelli05}.

\subsection{Results of the morphological analysis }
\label{morphres}

\begin{figure*}
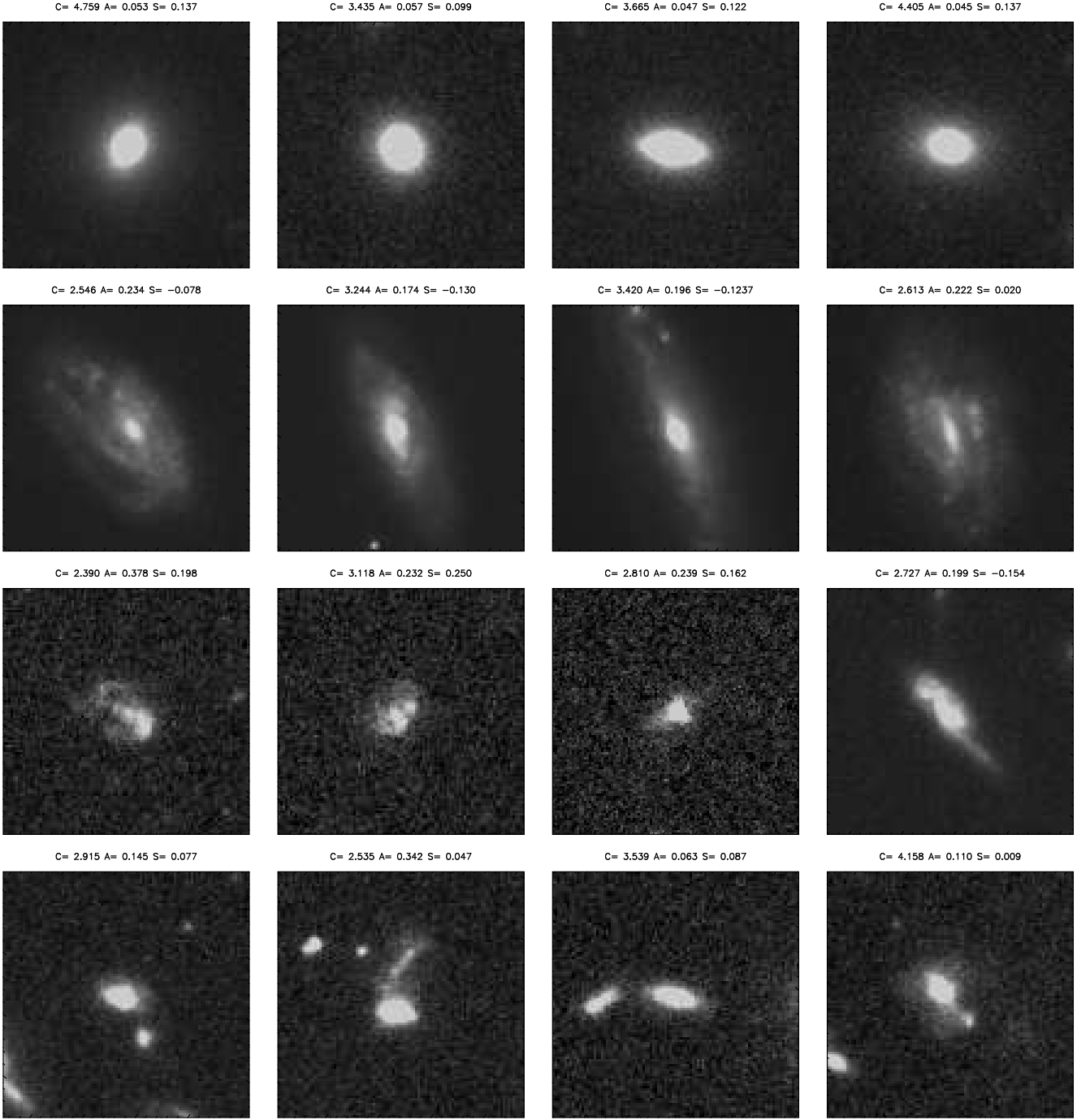

\centering
\begin{tabular}{c c c c}
\rotatebox{270}{\includegraphics[height=.23\textwidth]{103.ps}}      &
\rotatebox{270}{\includegraphics[height=.23\textwidth]{191.ps}}      &
\rotatebox{270}{\includegraphics[height=.23\textwidth]{133.ps}}      &
\rotatebox{270}{\includegraphics[height=.23\textwidth]{144.ps}}     \\
\rotatebox{270}{\includegraphics[height=.23\textwidth]{116.ps}}      &
\rotatebox{270}{\includegraphics[height=.23\textwidth]{118.ps}}      &
\rotatebox{270}{\includegraphics[height=.23\textwidth]{127.ps}}      &
\rotatebox{270}{\includegraphics[height=.23\textwidth]{156.ps}}     \\
\rotatebox{270}{\includegraphics[height=.23\textwidth]{146.ps}}      &
\rotatebox{270}{\includegraphics[height=.23\textwidth]{147.ps}}      &
\rotatebox{270}{\includegraphics[height=.23\textwidth]{160.ps}}      &
\rotatebox{270}{\includegraphics[height=.23\textwidth]{173.ps}}     \\
\rotatebox{270}{\includegraphics[height=.23\textwidth]{179.ps}}      &
\rotatebox{270}{\includegraphics[height=.23\textwidth]{3.ps}}      &
\rotatebox{270}{\includegraphics[height=.23\textwidth]{22.ps}}      &
\rotatebox{270}{\includegraphics[height=.23\textwidth]{75.ps}}     \\
\end{tabular}
\caption{Examples of objects visually classified as ellipticals, spirals, irregulars and
pairs (from upper to lower row) and their CAS parameters. The size of the cutouts is of 
5 $\times$ 5 arcsec.}
\label{figviscas}
\end{figure*}

In the 10.56 sq. arcmin region of the ACS field uncontaminated by UGC 10214, 
there are 177 objects with a SWIRE counterpart. As a first step, a visual inspection
of these sources was carried out on the I-band image
by three of the authors, independently. The individual
results were then compared and finally combined after the very few dubious cases were 
discussed (the excellent quality of the ACS data allows for very little doubt about
the morphological characteristics of the objects at these relatively bright magnitudes). 
The visual analysis identifies four main groups of objects:
$spheroids$ (consisting of ellipticals and S0 galaxies), $spirals$, 
$irregular$ galaxies and $pairs$ and found four objects
too faint to be classified; 21 were stars (stars were double-checked adopting the criterion of
\citealt{benitez04} for stellar classification, i.e. a SExtractor CLASS\_STAR $\ge 0.94$); 
37 spheroids; 69 spiral galaxies; 
24 irregulars and 22 merger or interacting systems or sources with multiple components
(all characterised as $pairs$). Fig. \ref{figviscas} shows examples of each of the categories
along with their CAS parameters, discussed hereafter.

For nearly a quarter of the 156 galaxies (the 21 stars are from now on excluded), 
CAS parameters were not computed.
Before computing CAS, pixels below 2$ \times$ RMS are
filtered out (corresponding to 25 mag/arcsec$^2$). For objects with no pixels above this level
the $A$ and $S$ are not computed. The computation of $C$ makes use of the SExtractor-derived
growth curve of each object and no filter is applied on the image.
For objects with complex structure (visually classified as pairs or irregular 
galaxies), their multiple components or uncommon 
shape render the CAS analysis difficult. For the
rest of the sources, however, this approach gave results largely in agreement with the visual
inspection. Spheroids tend to have lower $A$ and higher $C$ than spirals, irregulars or
pairs and tend to concentrate around $S \sim 0.1$. In fact, 36 of the 37 visually 
classified spheroids lie in or at the very borders of the region shown in 
Fig. \ref{figcasswire}, defined by:

\begin{figure*}
\centerline{
\psfig{file=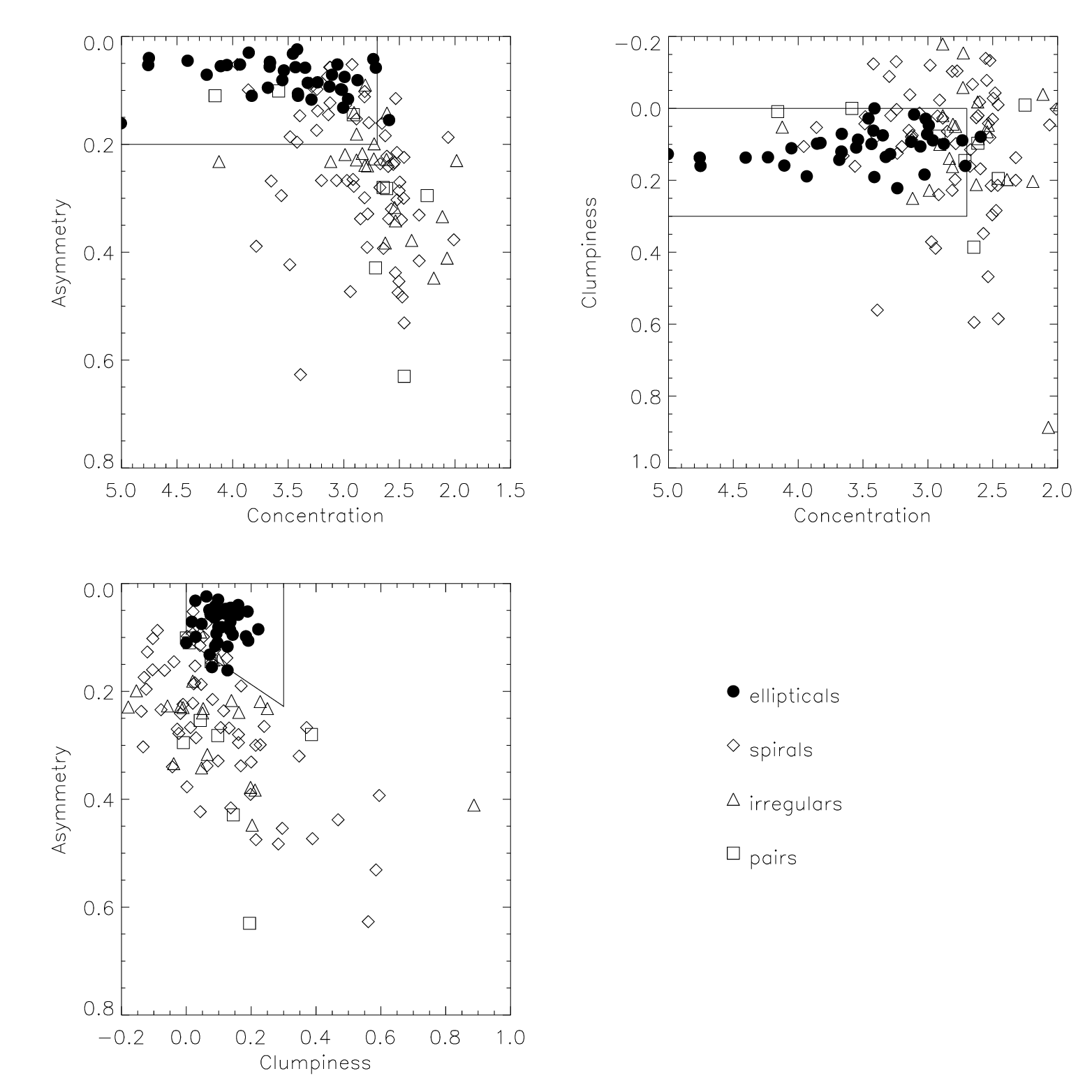,width=17cm}}
\caption{Asymmetry, Concentration and Clumpiness for the SWIRE-ACS galaxies and
comparison with the classification obtained by visual inspection.}
\label{figcasswire}
\end{figure*}

\begin{equation}
A < 0.2 \, \& \, C > 2.7
\label{eqAC}
\end{equation}
\begin{equation}
C > 2.7 \, \& \, 0.0 \le S < 0.3
\label{eqCS}
\end{equation}
\begin{equation}
0.0 \le S < 0.3 \, \& \, 3A - S < 0.3
\label{eqAS}
\end{equation}

\begin{figure*}
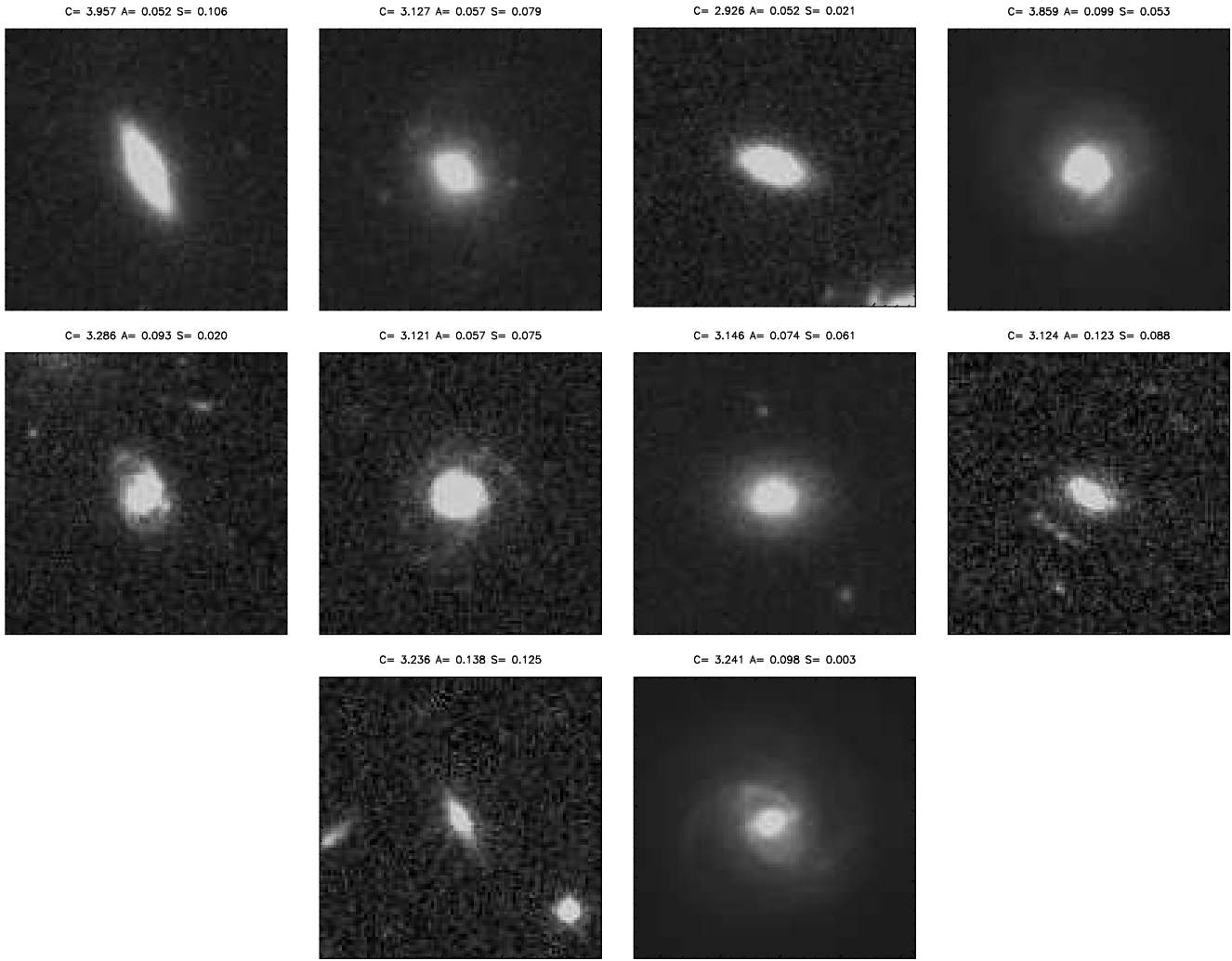

\centering
\begin{tabular}{c c c c}
\rotatebox{270}{\includegraphics[height=.23\textwidth]{30.ps}}      &
\rotatebox{270}{\includegraphics[height=.23\textwidth]{143.ps}}      &
\rotatebox{270}{\includegraphics[height=.23\textwidth]{66.ps}}      &
\rotatebox{270}{\includegraphics[height=.23\textwidth]{78.ps}}     \\
\rotatebox{270}{\includegraphics[height=.23\textwidth]{32.ps}}      &
\rotatebox{270}{\includegraphics[height=.23\textwidth]{170.ps}}      &
\rotatebox{270}{\includegraphics[height=.23\textwidth]{246.ps}}      &
\rotatebox{270}{\includegraphics[height=.23\textwidth]{234.ps}}      \\
 &
\rotatebox{270}{\includegraphics[height=.23\textwidth]{224.ps}}     &
\rotatebox{270}{\includegraphics[height=.23\textwidth]{205.ps}}      &
 \\
\end{tabular}
\caption{The ten galaxies visually classified as spirals that lie in the $CAS$ spheroid
region.}
\label{fig10obj}
\end{figure*}

An additional set of 10 objects fall within this same region.
Those objects are spirals but at least seven of them have prominent bulge components, as seen
on the cutouts illustrated in Fig. \ref{fig10obj}. Note that for some of them it is difficult to
see the spiral arms or disk components on the cutouts due to the chosen contrast 
but they are clear enough on the actual image for the objects to be classified as spirals.
We therefore deduce that more than 93\%
of the objects in the designated area of the CAS parameter space are ellipticals or 
bulge-dominated spirals (lower limit if one considers the 36 ellipticals and
the seven clearly bulge-dominated spirals among the 46 objects lying in this area).

This analysis sets the lower and upper limits of the number of spheroids in 37 (visually classified) 
and 46 (included in the region delineated by Eqs. \ref{eqAC}-\ref{eqAS}), 
respectively, within the $\sim 10$ sq. arcmin of effective area of the UGC 10214 ACS field.

\begin{figure*}
\centerline{
\psfig{file=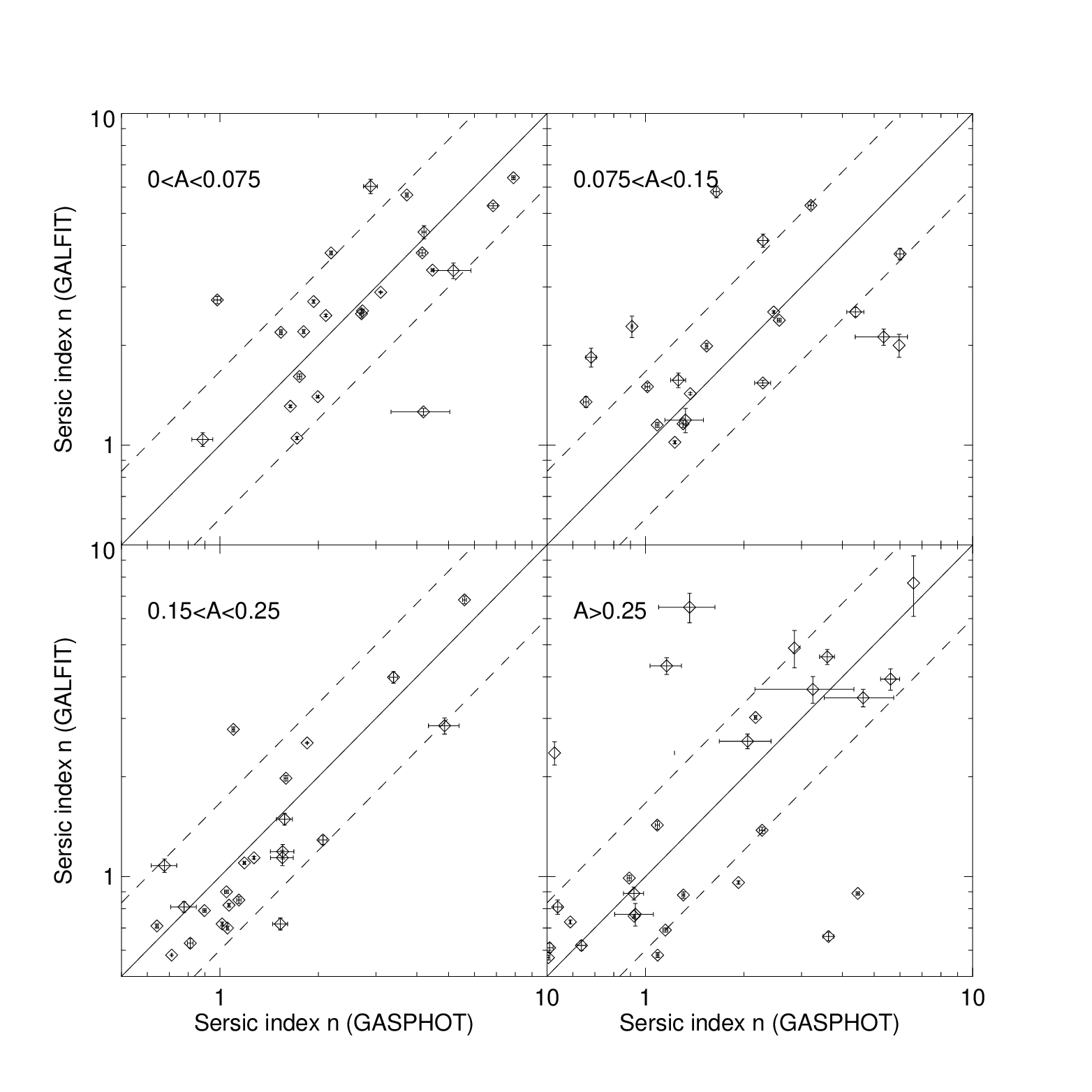,width=17cm}}
\caption{S\'ersic index, $n$, computed by GALFIT versus $n$ computed by GASPHOT for the
SWIRE-ACS galaxies, in bins of Asymmetry. The two tools provide comparable results for
low asymmetry objects. The dashed lines mark the region where $\Delta n/\bar{n} \le 0.5$,
occupied by $\sim$ 80\% of the objects. Objects with $n < 0.5$ - 22 in total -  
or objects for which one of the methods did not converge are not shown.}
\label{figmorphcompasym}
\end{figure*}

\begin{figure}
\centerline{
\psfig{file=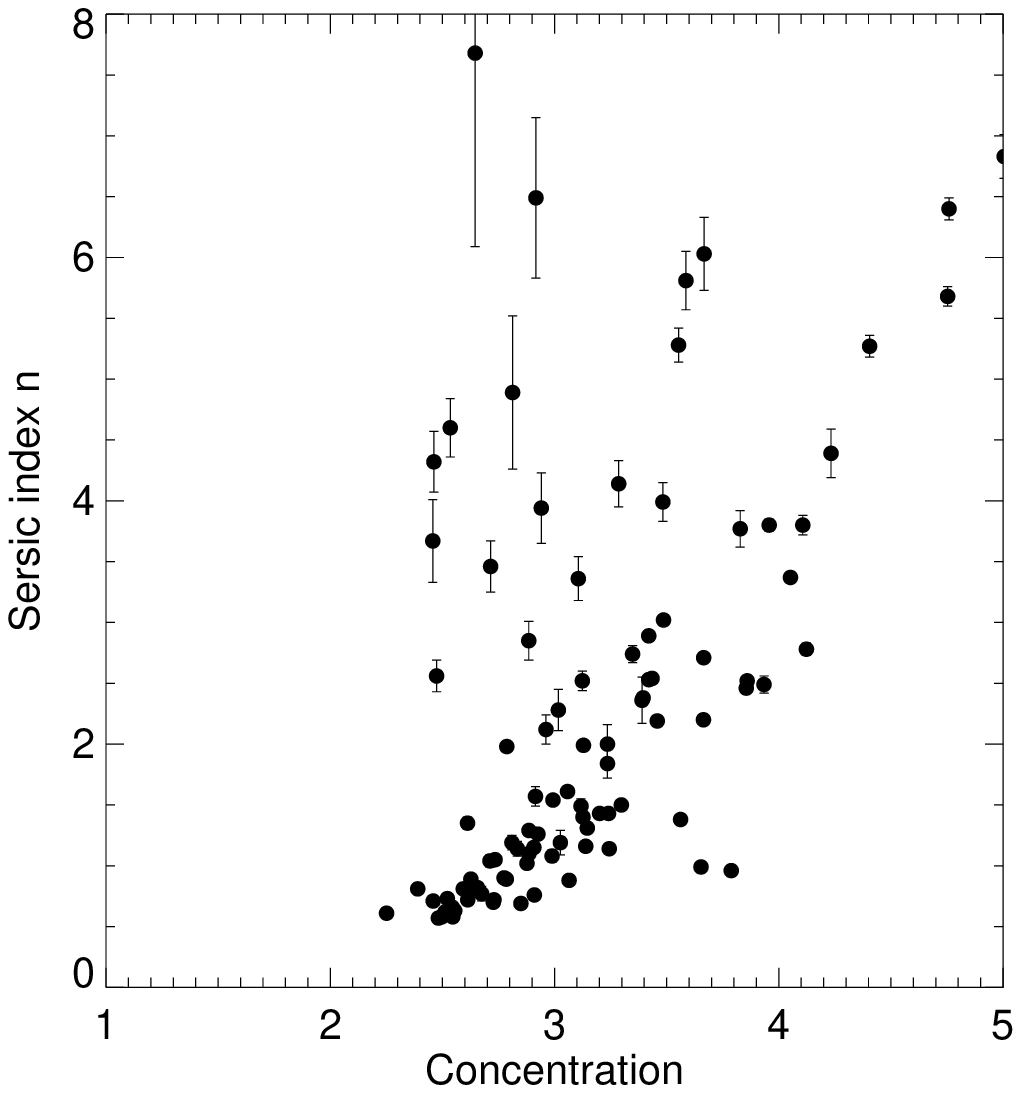,width=8cm}}
\caption{S\'ersic index, $n$, computed by GALFIT versus Concentration, $C$. 
$n$ scales with $C$ in the majority of cases. The objects with the largest scatter are
also those with the largest error bars and, in the majority of cases, have been
visually classified as pairs (including systems with various components).}
\label{fignvsc}
\end{figure}

The results of the parametric methods are more difficult to interpret. 
For low asymmetry objects the results of GALFIT and GASPHOT are in reasonable
agreement with each other (Fig. \ref{figmorphcompasym}) but discrepancies arise
for the objects with with $A > 0.25$. This is somewhat expected as the less symmetric
galaxies can not be easily reproduced by parametric models that are intrinsically 
symmetric (see \citealt{cassata05} and \citealt{pignatelli05} for further details 
on the comparison between GASPHOT and GALFIT on real and simulated galaxies, respectively). 
Some 80\% of the objects have $\Delta n/\langle n \rangle$ less than 0.5, region
indicated by the dashed lines on Fig. \ref{figmorphcompasym}.
Note that for some of the objects the two codes failed to converge
and therefore do not appear on this plot (31 for GALFIT and
25 for GASPHOT, with 20 objects in common between the two failing groups).
Parametric and non parametric approaches
also agree up to a point, with the S\'ersic index, $n$, correlating with $C$ in the majority
of the cases (Fig. \ref{fignvsc}).  Some 15 sources, however, present large deviations.
For these objects the errors on the estimated values of $n$ are
very large due to their nature: they are all pairs, sources with multiple components
or (just one case) objects with very small isophotal area.

The distribution of S\'ersic index, $n$, given by GALFIT is roughly bimodal, 
culminating at a value around 1 for late-type and around 2 for early-type 
galaxies. However, one third of the 
galaxies visually classified as early types have an $n$ lower than 2, and about 
the same fraction of late-type galaxies have $n$ greater than 2, implying that 
using a classification criterion based on the S\'ersic index alone
(see e.g. \citealt{ravindranath04}) would result 
in a certain amount of mis-classifications. Among the early type galaxies with 
a low value of the S\'ersic index, a large fraction are S0 galaxies while a 
certain number have a too small isophotal area to get reliable fits of the 
surface brightness distribution. Finally, the largest part of late-type galaxies 
having large S\'ersic indexes is mostly composed by bulge dominated systems or 
again objects occupying too small isophotal areas. Note that there was no lower 
limit imposed on the isophotal area in order to perform the fits, however all 
objects for which the fit failed to converge consisted of less than 50 pixels. 

The entire morphological analysis may be subject to an, unaccounted for, morphological
K-correction. Morphological classification performed at a given band suffers from this effect
as galaxies of higher redshifts are actually observed at bluer restframe wavelengths
(\citealt{windhorst02}; \citealt{papovich03}; \citealt{cassata05}).
\cite{windhorst02} found that the morphology of galaxies changes when moving from
the optical to the UV passbands. However, our analysis should not be
much affected by this effect as we do not expect a large number of high redshift galaxies,
due both to the relatively bright fluxes and to the very small size of the field. 
This, in fact, somewhat insures that we study galaxies with restframe between B- and I- bands. 
In particular, the number of high-redshift spheroids, population we are mostly interested in,
must be particularly low. \cite{rowan05}, based on photometric redshift estimations
and model predictions, showed that no more than 25\% of the 3.6 \mums galaxy sample 
down to a limit of 10 \textmu Jy have redshifts larger than $\sim$1, the largest fraction
of which are late-type objects.
CAS analysis may also suffer from the effects of the morphological K-correction
(see i.e. \citealt{conselice00}; \citealt{lotz04}), as in general, moving toward
shorter wavelengths results in larger values of $A$ and $S$ and smaller $C$. However,
since we ignore the real redshifts of our objects and decided not to rely upon the
photometric redshifts provided by \cite{benitez04} -- calculated based on three optical
passbands only -- in order to avoid introducing further uncertainties, we chose not to apply any kind
of correction.

Finally, the effects of the aperture on the estimation of the CAS parameters have not
been considered here. The parameter most sensitive to the size of the aperture is $A$ 
\citep{conselice00} and is more likely to be affected in the 
case of faint, high-redshift objects. However, since much of our analysis is based on the $C$
parameter and since the vast majority of our objects have very bright ACS 
counterparts (see Fig. \ref{figInumbercounts}, left panel), we are confident that 
aperture effects do not bias our work. The fact that visual inspection and CAS analysis 
are in such a good agreement when selecting early-type objects is also in support of
our argument.

\section{GALAXY NUMBER COUNTS }
\label{numbers}

\subsection{Observed number counts}
\label{coun}

\begin{figure*}
\centerline{
\psfig{file=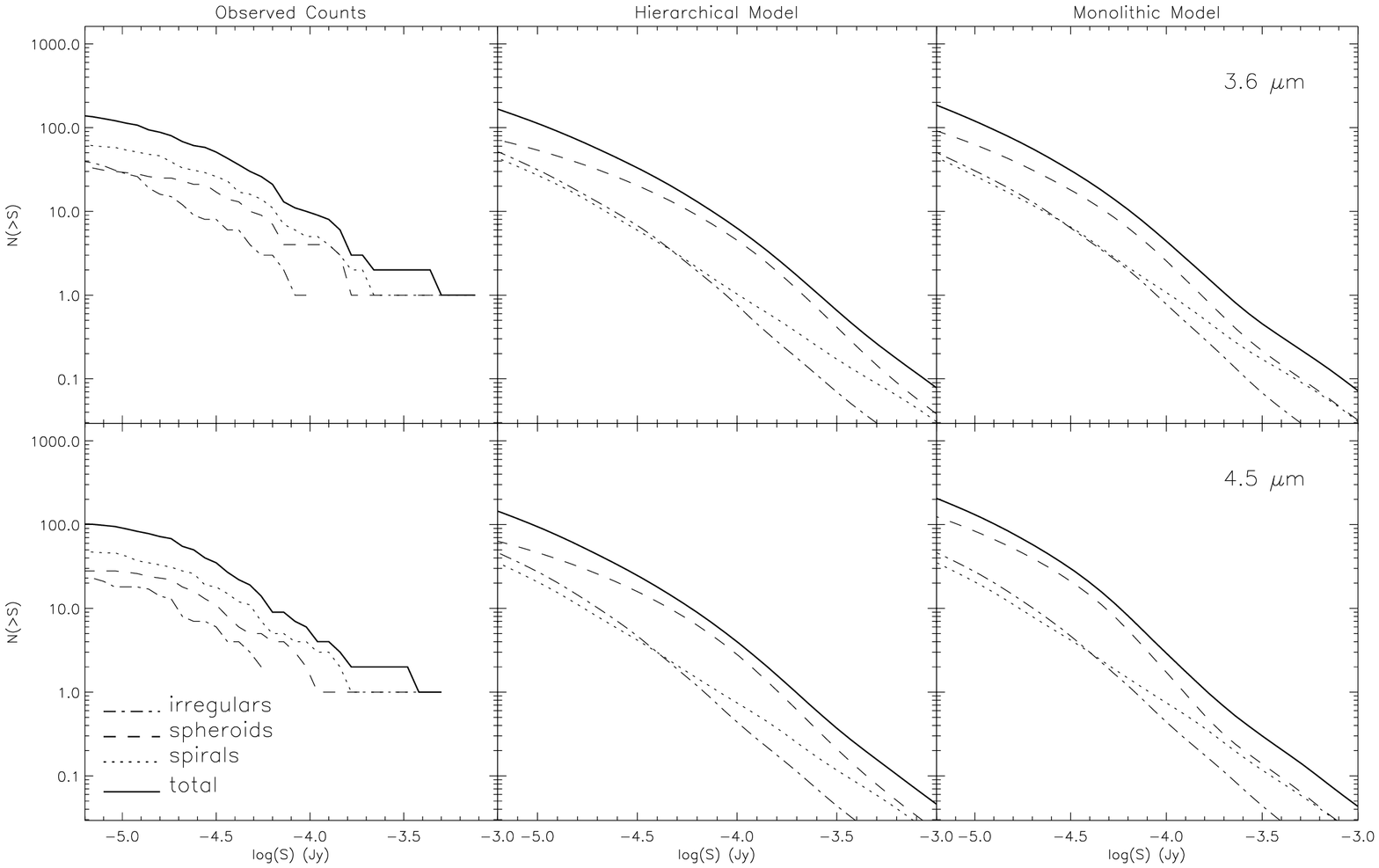,height=19cm,angle=90}}
\caption{Left column: observed cumulative 3.6 (upper panel) and 4.5 \mums (lower panel) counts
for the various types of objects (as classified by visual inspection) and their total
number (solid line). Middle and right columns: cumulative model counts for the $hierarchical$ (middle) and
$monolothic$ (right) models. The counts are reported here for the various
different morphological classes as indicated in the line caption. Note that $monolithic$
and $hierarchical$ scenaria only differ in the spheroids' (and therefore also total)
counts.}
\label{fignumbercounts}
\end{figure*}

The IRAC1 and IRAC2 channel (3.6 and 4.5 \mum) observed cumulative galaxy counts
are reported in 
the left column of Fig. \ref{fignumbercounts}.
The various morphological classes are these derived from the visual inspection.
We now confine our analysis to fluxes brighter than 
10 \textmu Jy in the two bands, and assume that the sample is 90\% complete here. 
This flux-limited sample consists of a total of 109 galaxies among which 28 are spheroids.

It is interesting to note here that the IRAC channels 1 \& 2 are optimally suited 
for the identification and analysis of spheroidal galaxy populations. 
We have performed an automated analysis of the I-band images over entire ACS 
field using the CAS parameter set and found that only $\sim$ 7\% of the galaxies to the 
survey limit I=27.2 lie in the area defined by Eqs. \ref{eqAC} -- \ref{eqAS}, and are therefore
classifiable as spheroids. The same analysis performed 
on the counterparts of the SWIRE IRAC population
shows a much larger incidence ($\sim$ 20\%) of early-type galaxies.
This is the result of the IR selection, favouring the detection of galaxies with old
stellar populations. Deeper SWIRE observations might reveal later type IR counterparts
of the ACS sources, however this is not evident from the actual distribution of 
morphologies of the IRAC sources with the I-band magnitude: the percentage of
early-type objects in bins of magnitude in the interval $[19.0, 25.0]$ 
takes values between 15\% to 45\% but without demonstrating any specific trend.

\subsection{Model number counts}
\label{model}

We now compare the statistical observables derived from our Spitzer IRAC dataset 
with  simple evolutionary models. While the details on the models are more extensively 
described in Rodighiero et al. (2005, in prep.), we provide here a summary of our 
general approach. We adopt in the following $H_0$=70, $\Omega_M$=0.3, $\Omega_{\Lambda}=0.7$.

In our approach four main galaxy classes are considered:
spheroids, quiescent spirals, an evolving population of irregular/merger
systems (hereafter starburst population), and minor contributions from AGN.

The SED templates describing the spectral shapes at different
galactic ages, needed to calculate the K-corrections,
have been computed using the stellar population synthesis code GRASIL
\citep{silva98}. We adopt a Salpeter IMF with a lower limit $M_l=0.15 M_{\odot}$
and a Schmidt-type law for the star formation (SF) rate:
$\Psi(t)=\nu M_g(t)$, where $\nu$ is the efficiency and $M_g(t)$ is
the residual mass of gas.  A further relevant parameter is the
timescale $t_{infall}$ for the infall of primordial gas. The evolution
patterns for the models considered here are obtained with the
following choices of the parameters.
For early-types: $t_{infall}=1~Gyr$, $\nu=1.3~Gyr^{-1}$; for
late-types: $t_{infall}=4~Gyr$, $\nu=0.6~Gyr^{-1}$.
The corresponding SF law for ellipticals have a maximum at
galactic ages of 1.4 Gyr, and is truncated at
3 Gyr to mimic the onset of a galactic wind.
For late-type galaxies, the peak of the SF occurs at 3 Gyr.
The parameters assumed to reproduce the spectra of spirals
and irregular galaxies are that of a typical Sb spiral.
This may not be very representative of an evolving population
during a starburst phase, however, given the spectral region considered
in this work, our assumption is still a good approximation.
We have then generated two grids of model SEDs for both early- and late-types,
spanning a range of ages from 0.1 to 15 Gyr.
For what concerns the spheroids, we made the assumption that these stellar systems 
are gas and dust free, so that extinction is negligile. Extinction is instead considered for 
late-type galaxies, with an evolution of the dust to gas ratio typical of an 
Sb galaxy (we assume the parameters proposed by \cite{silva98}).
Heated dust possibly contributes to the emission from the late-type
galaxies at rest-frame wavelengths longer than 2-3 \mum, which could
effect the derived model flux densities in the IRAC bands.
This will be considered in a future work.

For the local luminosity function we have made use of that estimated by 
\cite{kochanek01} for both early-type and late-type galaxy classes and derived from 
a K-band selected sample taken from the Two Micron All Sky Survey (2MASS), including 
4192 low-redshift ($z \sim 0.02$) galaxies.

Our schematic model for both the spheroidal and the spiral population 
assumes that the galaxy comoving number density keeps constant once the population 
is formed at a given redshift, while the galaxy luminosities evolve following their 
evolutionary stellar content.

On the other hand, the analysis of deep galaxy surveys in the K-band 
(\citealt{cassata05}; Franceschini et al. 2005 in prep.) indicates the presence of a 
numerous population of irregular/merging systems at high-redshifts, likely requiring 
luminosity as well as density evolution going back in cosmic time. We then add a 
population of starbursts whose density $\rho(z)$ evolves with the following rule:
\begin{equation}
\rho(z) \propto \rho(z_0) \times (1+z) 
\end{equation}
for $z < 1$, and constant above, and whose luminosities $L(z)$ increases as
\begin{equation}
L(z) \propto L(z_0) \times (1+z)^{1.3} 
\end{equation}
for $z < 2$, and constant above.

Before detailing the comparison of our model predictions
with the observed statistics derived from our sample, we
need to consider how a morphological selection correlates
with a classification based on the intrinsic SED of galaxies.
A convenient way to directly check this point is to look
at the colours of the SEDs templates assumed by the model,
in order to verify if they are representative of the
morphologically selected spheroids and spiral populations.

\begin{figure}
\centerline{
\psfig{file=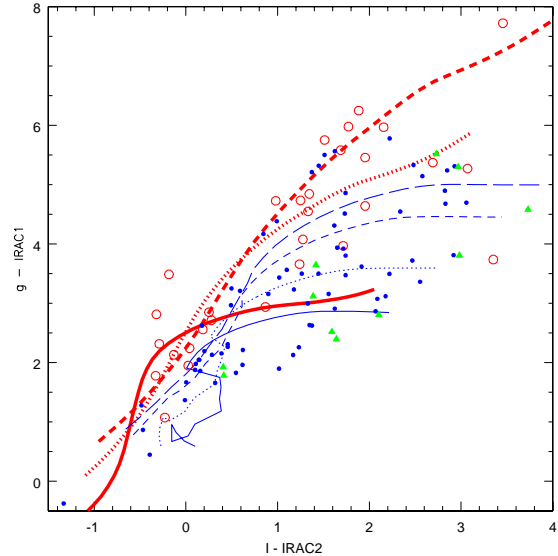,height=8cm}}
\caption{Colour-colour diagram (g-IRAC1) vs (I-IRAC2) showing the location
of the spheroids, spirals and pairs (in open circles, dots and filled triangles,
respectively) and the model evolutionary tracks of ellipticals and spirals of various
ages (in thick red and thin blue lines, respectively; 
for details about the line coding, see text).}
\label{figcolcol}
\end{figure}

As an example, in Fig. \ref{figcolcol} we report the I-4.5 \mums versus
the  g- 3.6 \mums observed colour-colour plot, compared
with the evolutionary tracks derived using model templates.
The blue dots mark the visually classified late-type sources;
red open circles are spheroids; green triangles are galaxy pairs.
The curves show the evolutionary tracks corresponding to
elliptical and spiral SEDs at different ages (thick red solid - dotted - dashed
lines: ellipticals of 1, 2 and 10 Gyr, respectively; blue solid - dotted - dashed - 
long dashed: 3, 5, 10 and 15 Gyr, respectively), spanning the redshift range 
between 0 and 1.8.
Fig. \ref{figcolcol} suggests that morphologically selected elliptical
galaxies preferentially populate the reddest region of the observed
colour distribution. However, the colour degeneracy between
age and extinction prevents us from having an unambiguous
photometric selection for spheroids. This can also be seem on this figure, 
where spheroids and late-type galaxies occupy
contiguous and partially overlapping regions of the colour-colour space.
The above mentioned degeneracy between age and extinction have thus produced 
star-forming 
sources with observed red colours (generally attributed to old stellar systems).
Similarly, those spheroids that have undergone recent episodes of star
formation have bluer colours and then fall within the region
preferentially populated by late-types.

This analysis suggests that a pure morphological selection
is not obviously comparable with a colour criterion, and that
a given amount of contamination affects such kind of comparison.
However, assuming that the template SEDs discussed above are
reasonably representative of the morphological classes, we
have investigated the statistical properties of the SWIRE sample
and compared them with the model predictions.
Any conclusion based on this analysis has to be taken with
caution, in particular for the specific properties of the
different spheroidal and spiral/irregular classes.

We now explore two schemes for the formation of spheroidal galaxies.
The first one corresponds to the classic -- $monolithic$ -- scenario 
assuming a single impulsive episode for the formation/assembly of the 
field ellipticals, occurring at a high redshift ($z_{form} > 2.5$, e.g. 
\citealt{daddi00} and \citealt{cimatti02}).
In our current implementation, we assume a redshift of formation $z_{form}=3.0$.
In this single burst model the birth of the stellar populations is coeval to the formation
of the spheroids.

The second model for spheroidal galaxies -- the $hierarchical$ model -- describes 
a situation in which massive ellipticals form at lower redshifts 
through the merging of smaller units down to recent epochs. In such a case 
the formation of early-type galaxies is spread in cosmic time. 
Very schematically, we achieve this by splitting the local spheroidal galaxies into 
several sub-populations, each one forming at different redshifts (in coincidence with
the birth of a new populations of stars building up the younger spheroids).
All sub-populations 
have the same mass function but different normalizations, whose total at $z=0$ has to 
reproduce the local luminosity function.
This model is being tested against deep galaxy surveys in the K-band, showing a general
tight agreement between predictions for the different morphological classes and observations
(K20 \citealt{cimatti02}, HDFS \citealt{franceschini98} and \citealt{rodighiero01};
GDDS \citealt{abraham04}).
In our current implementation, the bulk ($\sim 60\%$) of the early-type mass function
is dynamically assembled in the redshift interval $1.1 < z < 1.6$, with additional
fractions being assembled at higher and lower $z$.
At this stage, and for simplicity, we do not consider extinction effects during the 
star formation phase of the spheroids.
We defer to Franceschini et al. (2005) and Rodighiero et al. 
(2005) for a more detailed discussion and additional applications of this modeling.

Fig. 
\ref{fignumbercounts} shows the predicted cumulative number counts at 
for the hierarchical (middle panel) and monolithic (right panel) 
scenario, respectively. The difference is 
essentially in the spheroidal population. 
The main value of these models is to provide, particularly for the spheroidal and normal 
spiral components, the simplest possible extrapolations of the local luminosity 
functions back in cosmic time. The addition of a strongly evolving population of 
irregular/mergers is instead motivated by the need to reproduce, at least at the 
zero-th order, the observed statistics from deep K-band counts (Franceschini et al. 2005).
In any case, these simple models predict that spheroidal galaxies would make 
important contributions to the Spitzer/IRAC counts at fluxes above 10 \textmu Jy.

\subsection{Discussion}
\label{discuss}

\begin{figure*}
\centerline{
\psfig{file=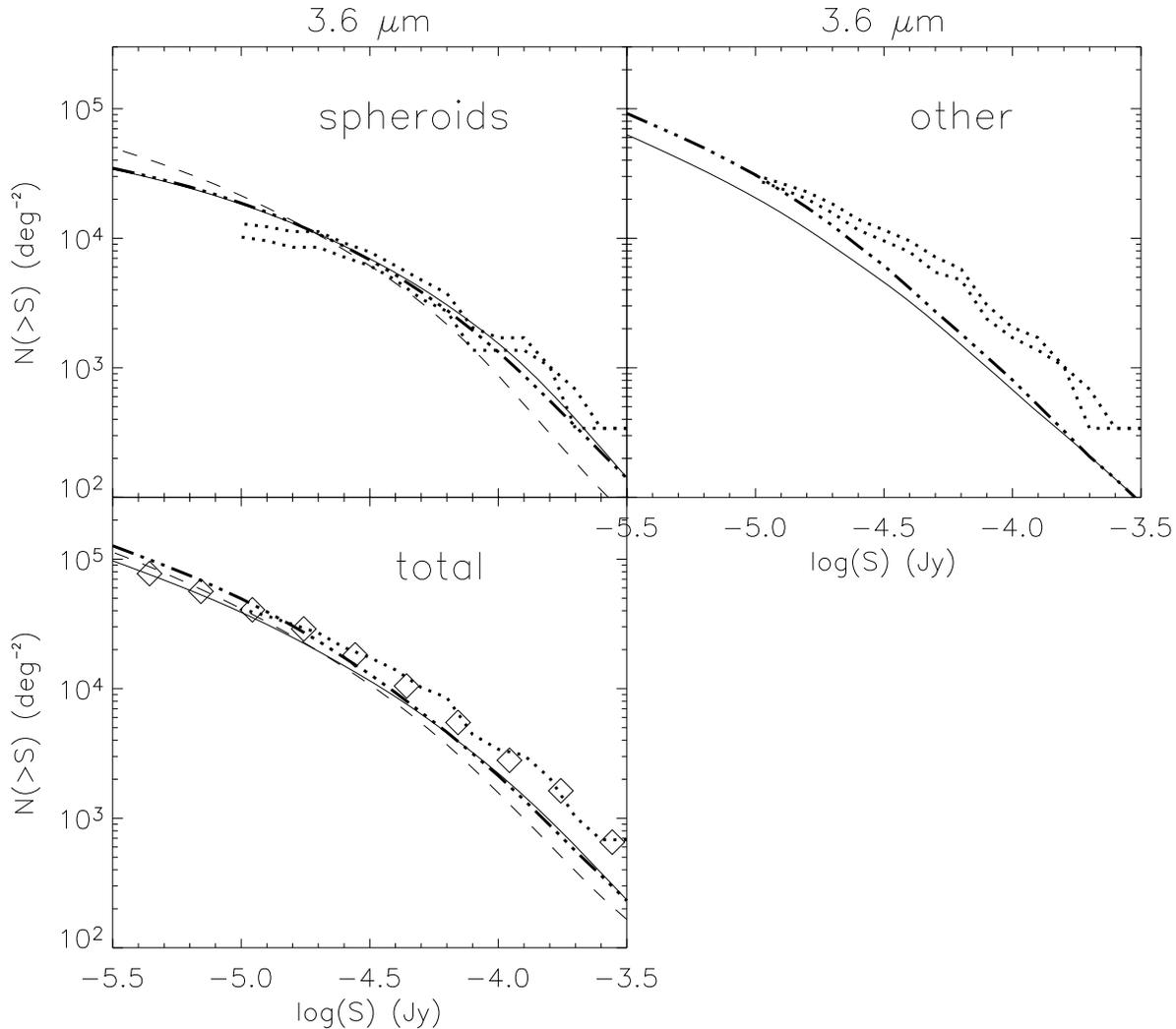}}
\caption{Comparison of the observed (dotted line) and model (solid and long-dashed 
lines for the hierarchical and monolithic scenario, respectively) 3.6 \mums number 
counts. The upper and lower limits for the observed counts are given by the CAS 
parameter analysis and visual inspection, respectively, as discussed in the text.
Note that, compared with Fig. \ref{fignumbercounts}, the spiral and irregular 
galaxies are here combined in the ``other" class. The diamonds in the lower panel
represent the IRAC1 counts given by Fazio et al. (2004). The dashed-triple-dot line
shows the predictions of an improved version of the model presented
in Xu et al. 2003 (see text for details).}
\label{fig36numbercounts}
\end{figure*}

\begin{figure*}
\centerline{
\psfig{file=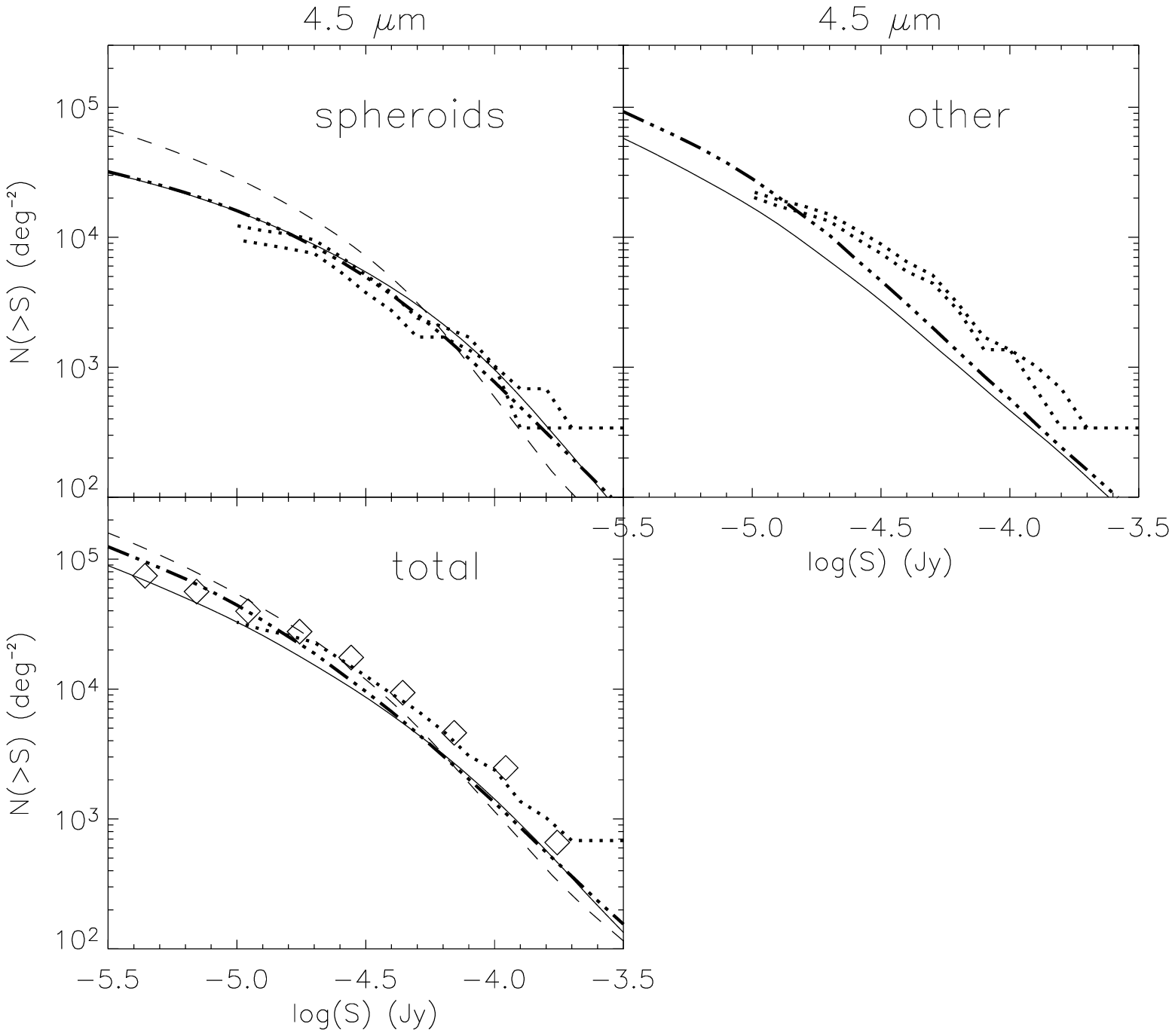}}
\caption{Same as Fig. \ref{fig36numbercounts} but for 4.5 \mum.}
\label{fig45numbercounts}
\end{figure*}

We report in Figs. \ref{fig36numbercounts} and \ref{fig45numbercounts} detailed
comparisons, with model expectations, of the SWIRE observed integral counts at 3.6 
and 4.5 \mums for the various morphological classes in the {\sl Tadpole} ACS image.
Only a general distinction between spheroidal and non-spheroidal 
systems is made here, the latter including both spirals and irregulars/mergers.
The observed counts are shown down to 10 \textmu Jy, roughly the 90\% completeness 
limits of SWIRE IRAC1 band.
The solid and dashed lines correspond to the hierarchical and monolithic scenario,
respectively, while the dotted lines show the observed number counts.
More precisely, the two dotted lines appearing in these figures
for each morphological class correspond to the lower and upper boundaries on the 
source real density which have been discussed in Sect. \ref{morphres}.
The upper and lower limits in this case of spheroids correspond to the CAS analysis
and visual inspection, respectively, whereas in the case of the remaining classes
(marked in the plot as ``other'') the upper limit is due to visual inspection and
the lower to the CAS parameters. 
The observed IRAC1 counts given by \cite{fazio04} are shown for comparison (marked as 
diamonds in Figs. \ref{fig36numbercounts} and \ref{fig45numbercounts}).
Note that, given the modest statistics of our source sample, the detailed shapes 
of the counts (showing bumps and humps at the bright end of the counts) 
are not to be considered as significant. 

Despite the very small size of our sample, the total counts are in good
agreement with those observed by \cite{fazio04} over a much larger area
(several square degrees), especially the
3.6 \mums counts. In the case of the 4.5 \mums counts, the statistics start being really poor 
as we are dealing with only 104 objects detected in this band down to the adopted flux limit.
The observed spheroidal counts are in very good agreement with those
predicted by the hierarchical scenario while the monolithic model suggests steeper counts.
However, while the models
predict counts essentially dominated by ellipticals (see 
middle and right panels of Fig. \ref{fignumbercounts}),
we observe an excess of late-type galaxies and merger-irregulars to dominate
the counts. Comparison between this same model and K-band observations did not show any
discrepancy.

The observed spheroids counts are inconsistent with the 
monolithic predictions that implies too steep number counts and would not only exceed 
the observed counts at the faint fluxes but also underpredict the bright end 3.6 \mums 
counts. A coeval origin for spheroids at high redshift (i.e. formation of
their stars in an istantaneous burst at $z>3$), 
therefore, seems to produce excess counts compared to observations. 
Changing the stellar IMF during the star-formation phase from a Salpeter to e.g. a 
\cite{scalo86} distribution would probably somehow ease the inconsistency \citep{cimatti02},
but at the cost of not explaining by large factors the extragalactic 
background radiation intensities 
(\citealt{madau00}, \citealt{franceschini01}) and the observed metalicities 
of intracluster plasmas \citep{mushotzky97}.

On the other hand, the observed numbers of spiral and irregular
galaxies are larger than that predicted by our simple evolutionary scheme. This
simple model, consistently fitting the K-band galaxy statistics, falls short of the observed
counts at both 3.6 and 4.5 \mum. The rising of the contribution of the
late-type population to the 3.6 and 4.5 \mums emission, with respect to its contribution at
2.2 \mum, might suggest a much stronger evolution (in both density and luminosity)
for star-forming galaxies at longer wavelengths. Our results might
indicate that the 3.6 \mums band selects more active galaxies than the K-band (2.2 \mum).
If in the model the starburst and spiral populations are then assumed to contribute
more significantly to the source counts, the discrepancy with observations should
be reduced. Moreover, a more suited set of SED templates describing the star-forming
population should be included (accounting for PAH emission and a more
accurate description of dust and extinction, given that in our approach
we roughly assume similar spectra for spirals and evolving starburst).
However, larger sample and statistics are needed to check this
assumption.

Very qualitatively, it might be tempting to interpret the observed excess population 
of irregulars and mergers and their inferred strong cosmological evolution as the 
progenitors of spheroidal galaxies, the rapid relaxation and gas exhaustion following 
the merger bringing to a sudden appearance of a spheroid. Obviously, any more 
quantitative conclusion will require much further information on source 
identifications and redshifts.

Because of the inconsistency between observed and model late-type galaxies, we
further compared our results with an improved version of the model presented 
by \cite{xu03}. 
In its original version this model over-predicted the 24 \mums counts at flux
levels of $\sim$ 1 mJy (for a
detailed discussion see \citealt{papovich04} and Shupe et al. in preparation).
In order to correct for this effect but still fit the 15 \mums {\it ISO} counts
the evolution rate of starburst galaxies was lowered with respect to the old model
and the strength of the PAH features in the wavelength range of
5 \mums $< \lambda <$ 12 \mums was raised by a factor of 2 in the SEDs of both
starburst and normal galaxies. Thus, the predicted late-type counts are still below 
the observed limits but are higher by a factor of $\sim 1.5$ than those predicted 
by our simple hierarchical scenario. They are shown in
dashed-triple-dot line in Figs. \ref{fig36numbercounts} and \ref{fig45numbercounts}.
This reinforces our previous statement about the deficit of model late-type galaxies 
reflecting the lack of PAH emission from the models presented in Section \ref{model}.

\section{CONCLUSIONS}
\label{conclude}

This work presents results of a morphological analysis of a small subset of
the Spitzer Wide-area InfraRed Extragalactic survey (SWIRE) galaxy population.
Our sample is flux-limited at 3.6 and 4.5 \mums and consists of 156
galaxies. Our analysis
is based on public ACS data taken inside the SWIRE N1 field.
We distinguish two very general classes of galaxies, bulge- and disk- dominated ones,
the first class being referred at with the general term ``spheroids'' and the second
one containing everything from spirals to irregulars and pairs.
Even though the requirement for 3.6 and/or 4.5 \mums detections favours the 
selection of early-type galaxies, the sample under study is dominated by
a large fraction of disk galaxies and interacting systems ($\sim$ 80\%),
already suggesting that elliptical galaxies assemble late.

The $monolithic$ and $hierarchical$ models checked against this data set
considerably under-predict the number of late-type galaxies. The observed 3.6 and 
4.5 \mums early-type counts are in very good agreement with the estimations of
the hierarchical scenario, showing however a deficit toward the fainter end of the counts,
possibly reflecting some incompleteness that is already introduced at this flux level.
The monolithic predictions imply steeper counts and fail in reproducing the observations.
The disagreement is stronger at the faint end of the 4.5 \mums early-type counts.

Additional comparison of the dataset with another model available in the literature
\citep{xu03} suggested that dealing in an appropriate way with some of the models deficiencies
like the dust or PAH emission components should result in a better agreement.
It should be mentioned that the model predictions may depend significantly 
on the assumed stellar IMF (see e.g. \citealt{cimatti02}), although the one we adopted
here should be rather representative of the main phases of star formation.
The model itself is rough and the predicted counts should not be taken at face 
value but rather be considered indicative of the expected tendencies.

This ACS data set is the deepest
available, as of today, in any of the SWIRE fields and therefore the results of the
present analysis are of great importance for the understanding of the SWIRE galaxy
population. Due to the small size of our sample, however, we cannot be very conclusive
and can only outline some general tendencies. We would like to point out a number of 
potential sources of uncertainty in the 
interpretation of our results, namely: i) the way that large-scale structure may 
influence the results, given the particularly small size of the ACS field; 
ii) the possible existence of heavily obscured, by dust, ellipticals at the stage 
of their formation; iii) the assumption of {\it a priori} values instead of a 
$\chi^2$ minimization on some of the model parameters; and finally iv) the uncertainties on 
the exact shape of the SEDs of the various galaxy types in these newly explored, 
IR wavelengths. 

\vspace{0.75cm} \par\noindent
{\bf ACKNOWLEDGMENTS} \par
\noindent
This work is based on observations made with the {\it Spitzer Space Telescope},
which is operated by the Jet Propulsion Laboratory, California Institute of
Technology under NASA contract 1407.
Support for this work, part of the Spitzer Space Telescope Legacy Science
Program, was provided by NASA through an award issued by the Jet Propulsion
Laboratory, California Institute of Technology under NASA contract 1407.

ACS was developed under NASA contract NAS 5-32865, and this research
has been supported by NASA grant NAG5-7697. We are grateful for an
equipment grant from  Sun Microsystems, Inc.
The Space Telescope Science
Institute is operated by AURA Inc., under NASA contract NAS5-26555.

This work was supported in part by the Spanish Ministerio de
Ciencia y Tecnologia (Grants Nr. PB1998-0409-C02-01 and ESP2002-03716)
and by the EC network "POE'' (Grant Nr. HPRN-CT-2000-00138).

We would like to thank N. Benitez for making publicly available the reduced
ACS image used in this work.

Finally we thank the anonymous referee for the precise and detailed report that
greatly improved the presentation of our work.

\end{document}